\documentclass[aps,pra,twocolumn,superscriptaddress]{revtex4-2}
\usepackage[utf8]{inputenc}
\usepackage[T1]{fontenc}
\usepackage[english]{babel}
\usepackage{times,amssymb,mathtools,booktabs,array,graphicx,xcolor}
\usepackage[hidelinks]{hyperref}
\usepackage{xurl}
\newcommand{\ket}[1]{|#1\rangle}
\newcommand{\Id}{\mathbb I}
\definecolor{figplus}{HTML}{D0021B}
\definecolor{figminus}{HTML}{1550C8}
\definecolor{figzero}{HTML}{9AA5B1}
\begin{document}
\title{Geomagic Square Representation of a Real AME$(4, 6)$ State}
\author{\texorpdfstring{Wojciech Bruzda}{Wojciech Bruzda}}
\affiliation{Center for Quantum-Enabled Computing, Center for Theoretical Physics, Polish Academy of Sciences, Aleja Lotnik\'{o}w 32/46, 02-668 Warsaw, Poland}
\email{w.bruzda@cft.edu.pl}
\date{\today}
\begin{abstract}
Seeking a pictorial representation of four-party absolutely maximally entangled (AME) states of local size six, we independently construct a real AME$(4,6)$ state from a table of signed Pauli matrices. Its two-unitary matrix has $60$ nonzero entries, equal to $1$ or $\pm 1/\sqrt{2}$. A locally equivalent representative yields a genuine geomagic square, where every row, column and main diagonal tiles a hexagon and all $36$ pieces are pairwise non-congruent. Colors, shapes and hatching retain the full tensor, allowing its reconstruction from the picture.
\end{abstract}
\maketitle

\section{Pictorial motivation}\label{sec:motivation}

An absolutely maximally entangled state of four qudits, denoted AME$(4,d)$, is a pure state whose reduction to any two of its four $d$-level subsystems is maximally mixed. Equivalently, its rescaled coefficient tensor defines a two-unitary matrix of order $d^2$, whose flattenings for all three splittings of the four subsystems into two pairs are unitary~\cite{goyeneche}.

An AME$(4,d)$ state whose two-unitary matrix is a permutation encodes two orthogonal Latin squares. For example, at $d=3$ or $d=4$, the cell $(i,j)$ of a $d\times d$ array contains a regular $d$-gon divided into $d$ triangular sectors, so sector $k$ is filled with color $l$ whenever $T_{ijkl}=1$. Any two indices determine the other two, so the sectors of each row and column assemble the full polygon. This gives a semimagic analog of a geomagic square~\cite{sallows}, whose defining requirement also includes both main diagonals. All these single-sector pieces are congruent. This observation is the main motivation for the rest of this work.

For $d=6$ there is no permutation solution, since orthogonal Latin squares of order six do not exist~\cite{tarry,stinson}. Hence, Euler's problem of $36$ officers has no classical solution~\cite{euler}. Mutually orthogonal quantum Latin squares of order six without entanglement do not exist either~\cite{ball2026}. Complex AME$(4,6)$ constructions are known~\cite{rather,review}. A family of Butson-type~\cite{Butson}, two-unitary complex Hadamard matrices of order $36$, with entries proportional to $6^{\textrm{th}}$ roots of unity, was constructed in Ref.~\cite{bruzda2024}. The first order-$6$ perfect tensors constructed ``artisanally'' were obtained in Ref.~\cite{gross}, by applying Clifford unitaries separately to two-qutrit and three-qutrit subspaces. All previously known AME$(4,6)$ representatives have complex entries~\cite{rather,zyczkowski,review}, for example, in the matrix of Ref.~\cite{rather}, each row has two or four nonzero entries, with three moduli and phases given by powers of $e^{i\pi/10}$. 

In each case, the natural figure describing a possible pictorial representation is a hexagon. Its internal sub-areas can represent squared moduli but cannot record the phases responsible for interference. Hence, a faithful picture requires as few phases as possible, ideally real coefficients, whose signs require only a binary mark. Rational squared moduli are also convenient, since they permit exact geometric data.

\begin{figure*}[t]
\centering
\includegraphics[width=0.91\textwidth]{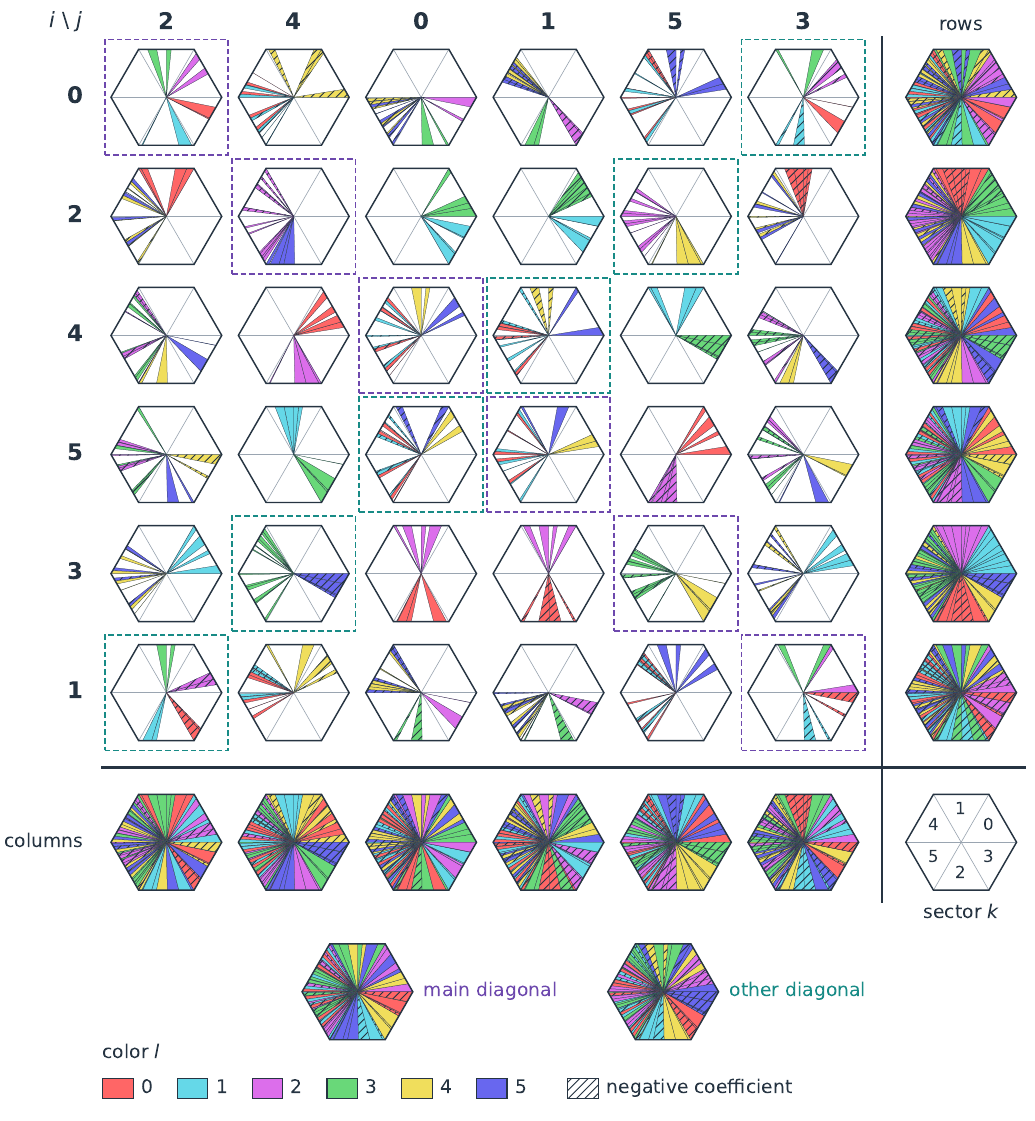}
\caption{One member of the genuine geomagic family encoding $T^{\textrm{g}}$ (\textrm{g} for geomagic), using the unequal interval splits described in Appendix~\ref{app:geometry}. Printed cell labels give $(i,j)$, the sector key gives $k$ and the color legend gives $l$. The total area of all fragments with fixed $(i,j,k,l)$ is $(T^{\textrm{g}}_{ijkl})^2$ in units of one triangular sector and hatching marks a negative coefficient. The right and bottom hexagons assemble each row and column. The two hexagons below assemble the main diagonals, whose cells have matching dashed frames. Colors and signs may differ between assemblies and their common target is the uncolored hexagon. Printed resolution limits measurements and hatch visibility. A color-independent version of this figure, with each color $l$ also encoded by a texture, is available in Ref.~\cite{github}.}
\label{fig:geomagic}
\end{figure*}

The present work originated from a search for such a representation, first manually, then assisted by a large language model (LLM). Besides exact pictures, the search unexpectedly produced a real AME$(4,6)$ state, relevant to the real-existence question raised in Refs.~\cite{rather,zyczkowski,review}. Its orthogonal two-unitary matrix has only $60$ nonzero entries, equal to $1$ or $\pm 1/\sqrt{2}$. This paper accompanies other papers that appear simultaneously or will follow shortly. Independent research projects arrived at the same answer. However, their motivations differ and the resulting matrices take visibly different forms. Their local unitary equivalence lies beyond the scope of this work. This work makes no claim of priority for the construction of a real AME$(4,6)$ state~\cite{private_SAR,private_SAR_GH,HZ26}. Its primary aim is the correspondence between the geometric representation and the internal structure of the state, in which areas, tilings, colors and hatching encode its coefficients and the normalization of its three flattenings. We provide the final tensor with a short proof of two-unitarity and construct the genuine geomagic square in Fig.~\ref{fig:geomagic}. The appendices contain the full matrix, drawing and reconstruction rules and a combinatorial check of non-congruence. Exact coordinates for the illustrated drawing, source files and the remaining technical material are collected in the GitHub repository~\cite{github}. The repository~\cite{github} also contains a companion note on self-similar geomagic representations of real AME$(4,6^n)$ states generated from the present geomagic square.

\section{Real AME$(4, 6)$ State}\label{sec:construction}

Let $i,j,k,l\in\{0,\ldots,5\}$ label the subsystems $A$, $B$, $C$ and $D$ and $U_{6i+j, 6k+l}=T_{ijkl}$, where
\begin{equation}\label{eq:state-matrix}
 \ket\psi=\frac16\sum_{i,j,k,l}T_{ijkl}\ket{i}_A\ket{j}_B\ket{k}_C\ket{l}_D.
\end{equation}
The following table defines $\Xi_x^{ab}$
\begin{equation}\label{pair-table}
\begin{array}{c|rrrr}
ab\backslash x&0&1&2&3\\\hline
AB&\textrm{Z}&i\textrm{Y}&\textrm{I}&\textrm{X}\\
AC&-\textrm{Z}&-\textrm{I}&\textrm{X}&-i\textrm{Y}\\
AD&i\textrm{Y}&\textrm{I}&-\textrm{Z}&\textrm{X}\\
BC&\textrm{X}&\textrm{Z}&\textrm{I}&i\textrm{Y}\\
BD&\textrm{I}&\textrm{Z}&-i\textrm{Y}&\textrm{X}\\
CD&\textrm{I}&\textrm{X}&\textrm{Z}&-i\textrm{Y}
\end{array},
\end{equation}
where the real matrices $\textrm{I}$, $\textrm{X}$, $\textrm{Z}$ and $i\textrm{Y}=\textrm{Z}\textrm{X}$ are the standard Pauli matrices. This should be understood as follows. For each pair $ab$ and each $x$, set the two named indices to $x$ and the remaining indices, in increasing subsystem order, to $4+u$ and $4+v$, with $u,v\in\{0,1\}$. Then assign the resulting entry the value $(\Xi_x^{ab})_{uv}/\sqrt2$. Additionally, set $T_q=1$ for each odd permutation $q$ of indices $(0,1,2,3)$. All other entries of $T$ vanish. Each table entry has two nonzero matrix elements. Thus $T$ has $60$ nonzero coefficients, namely $12$ equal to $1$ and $48$ to $\pm1/\sqrt2$. Since the sum of $T_{ijkl}^2$ equals to $12+48/2=36$, Eq.~\eqref{eq:state-matrix} is normalized. The full matrix and a worked example are in Appendix~\ref{app:construction-details}.

\section{Two-unitarity}\label{sec:two}

The other two flattenings are defined via reshuffling $U^{\textrm{R}}_{6i+k,\,6j+l}=T_{ijkl}$ and partial transpose $U^\Gamma_{6i+l,\,6k+j}=T_{ijkl}$, which are just particular rearrangements of the matrix entries.
In each cut, row and column permutations split the matrix into $\Id_{12}$, two $4\times4$ Bell-basis blocks and four $4\times4$ mixed blocks. The identity block emerges from the odd permutations, because each ordered pair of distinct levels in $\{0,1,2,3\}$ can be uniquely completed to odd permutation. The Bell blocks connect $(x,x)$ to $\{4,5\}^2$ and their rows or columns are the vectors $\operatorname{vec}(\Xi_x)/\sqrt2$ from a table row~\eqref{pair-table}, with transposes as required by the index order. They are orthonormal because $\textrm{I}$, $\textrm{X}$, $\textrm{Z}$ and $i\textrm{Y}$ are orthogonal with respect to the Hilbert--Schmidt inner product and each table row contains each of them exactly once, up to sign.

For mixed labels, one uses the order $(x4,x5,4x,5x)$ on both sides. Write each block as
\begin{equation}\label{eq:generic-block}
 K=\frac1{\sqrt2}\begin{bmatrix}P&Q\\R&S\end{bmatrix}.
\end{equation}
Its entries are
\begin{equation}\label{eq:three-mixed-blocks}
\renewcommand{\arraystretch}{1.3}
\begin{array}{l|cccc}
 &P&Q&R&S\\\hline
K_x&\Xi_x^{AC}&\Xi_x^{AD}&\Xi_x^{BC}&\Xi_x^{BD}\\
K_x^{\mathrm{R}}&\Xi_x^{AB}&(\Xi_x^{AD})^{\mathsf T}&\Xi_x^{BC}&\Xi_x^{CD}\\
K_x^\Gamma&(\Xi_x^{AC})^{\mathsf T}&(\Xi_x^{AB})^{\mathsf T}&\Xi_x^{CD}&\Xi_x^{BD}
\end{array}.
\end{equation}
All four constituents are orthogonal, so $KK^{\mathsf T}=\Id_4$ reduces to $PR^{\mathsf T}+QS^{\mathsf T}=0$. Substitution from Eq.~\eqref{pair-table} verifies this identity for every $x$ in all three cuts. For example, $K_3$ gives $-(i\textrm{Y})(i\textrm{Y})^{\mathsf T}+\textrm{X}\textrm{X}^{\mathsf T}=0$. Thus all blocks and all three flattenings are orthogonal. In consequence, every two-party reduction is $\Id_{36}/36$, which implies that $\ket{\psi}\in$ AME$(4,6)$.

\section{Genuine Geomagic Representation}\label{sec:geomagic}

Following the formal definition of Ref.~\cite{sallowsappendix}, a geomagic square of order $n\geqslant2$ is an $n\times n$ array of pairwise non-congruent compact planar pieces such that the $n$ pieces of every row, every column and both main diagonals tile one common target, which may be any compact planar set. Tiling means that congruent copies of the pieces, obtained by rotations, translations and reflections, have the target as their union and overlap only on their boundaries. We call such a square genuine to stress that, unlike the examples of Sec.~\ref{sec:motivation}, it satisfies both the diagonal and the non-congruence conditions.

The sparse tensor has twelve unit coefficients, each producing a congruent single-sector piece. To obtain distinct pieces, apply local orthogonal maps built from the Hadamard matrix $\textrm{H}=(\textrm{X}+\textrm{Z})/\sqrt2$, namely $O_A=\textrm{H}\oplus\Id_4$, $O_B=\textrm{H}\oplus\textrm{H}\oplus\Id_2$ and $O_C=\Id_4\oplus\textrm{H}\textrm{X}$:
\begin{equation}\label{eq:local-maps}
U^{\textrm{g}}=(O_A\otimes O_B)\,U\,(O_C\otimes\Id_6)^{\mathsf T},\quad T^{\textrm{g}}_{ijkl}=U^{\textrm{g}}_{6i+j,\,6k+l},
\end{equation}
where $\textrm{g}$ stands for geomagic. The transformed state remains AME$(4,6)$. The tensor $T^{\textrm{g}}$ has $152$ nonzero entries, with squares $\frac{1}{2}$, $\frac{1}{4}$ or $\frac{1}{8}$.

Let rows and columns be displayed in the order $(0,2,4,5,3,1)$ and $(2,4,0,1,5,3)$, respectively. Then, the main diagonals consist of $(i,D(i))$ and $(i,D'(i))$, where $D=(2,3,4,5,0,1)$ and $D'=(3,2,5,4,1,0)$. Reordering rows and columns leaves the tensor and the row and column tilings unchanged and only puts the cell sets $(i,D(i))$ and $(i,D'(i))$, for which Table~\ref{tab:transversals} is constructed, on the two geometric diagonals. This requires the rows paired by $D^{-1}D'=(0\,1)(2\,3)(4\,5)$ to occupy mirrored positions, which holds for $48$ row orders. Two-unitarity alone does not yield such a square. It only fixes areas, i.e., each piece has unit area and, in every row and column, the fragments in each sector add up to one full sector. It does not prevent overlaps, constrains no diagonal and does not exclude congruent pieces.

For the present tensor, Table~\ref{tab:transversals} supplies an exact decomposition
\begin{equation}\label{eq:geomagic-decomposition}
\begin{gathered}
(T^{\textrm{g}}_{ijkl})^2=\sum_\alpha w_{k\alpha}
\delta_{j,\tau_{k\alpha}(i)}\delta_{l,\sigma_{k\alpha}(i)},\\
\sum_\alpha w_{k\alpha}=1.
\end{gathered}
\end{equation}
Here $\tau$ and $\sigma$ are permutations and each $\tau$ agrees with $D$ and with $D'$ at exactly one $i$. Partition a sector's outer edge into intervals of total length $w_{k\alpha}$ for each transversal $\alpha$. Place the corresponding triangles, with apex at the hexagon center, in cells $(i,\tau_{k\alpha}(i))$ and paint them $\sigma_{k\alpha}(i)$. Every interval then occurs exactly once in every row, column and diagonal. Consequently all fourteen lines tile the full hexagon (only shared boundaries overlap). Each color also covers the hexagon once across all cells. Appendix~\ref{app:geometry} fixes their order and permits any positive split with the prescribed matching totals. Figure~\ref{fig:geomagic} shows one such choice. In this split the $36$ perimeters are pairwise different, which provides an additional metric check of non-congruence~\cite{github}. The combinatorial argument below holds for every split, including the canonical one of Appendix~\ref{app:geometry}.

Each piece has area $\sum_{k,l}(T^{\textrm{g}}_{ijkl})^2=1$ in sector units. Non-congruence follows from a finite combinatorial test. One can ignore colors and hatching and count the maximal segments covered by the piece on each of the six outer edges. The pieces contain between three and twelve lobes, meeting only at the center, which is their unique cut point, i.e., when removed, it disconnects the piece. Hence, a congruence must fix the center. All nonradial boundary segments lie on the edges of the hexagon, so the congruence maps edge lines to edge lines and it is a rotation or reflection in the symmetry group of the hexagon. Comparing these six counts up to the symmetry leaves only four candidate pairs. Appendix~\ref{app:geometry} distinguishes these pairs by lobe counts or by occupancy near a vertex. Thus there are no two pieces which are congruent, even under reflection, for any allowed positive interval split. Together with the fourteen tilings, this proves that the geomagic square is genuine. As allowed by the formal definition of geomagic squares, in which pieces are arbitrary compact sets~\cite{sallowsappendix}, the lobes retain their fixed relative positions and each closed piece is connected, although its interior is disconnected. As a result, the geomagic picture retains the complete real tensor, but the converse does not hold, since the picture is not unique. Other interval splits, sector placements, display orders, matching orders in Eq.~\eqref{eq:matching-words} or other exact decompositions replacing Table~\ref{tab:transversals} yield different drawings that all decode to the same $T^{\textrm{g}}$, while other locally equivalent representatives yield pictures of locally equivalent states. Areas encode magnitudes and hatching records the signs needed for interference. Appendix~\ref{app:geometry} provides the description of the inverse map.

\section{Outlook}\label{sec:outlook}

The geomagic picture turns the normalization conditions of the three flattenings of $T^{\textrm{g}}$ into areas, tilings and color balance and the orthogonality between different rows is introduced through the signs depicted by hatching. A drawing in which these cancellations become geometrically visible would complete the dictionary. Conversely, the areas of any geomagic square built from colored sectors define a nonnegative tensor and one may ask whether signs or phases attached to their square roots yield a two-unitary matrix. This tensor analog of unistochasticity may require phases already for magic squares. The squares of Ref.~\cite{sallows} offer natural test cases.

The drawing can also be optimized. Other locally equivalent representatives, sector placements or interval splits may reduce the $480$ triangular fragments of Fig.~\ref{fig:geomagic}, connect the interiors of the pieces or add symmetry. One can check the repository~\cite{github} for other possible attempts. Encoding phases (by a special direction of hatching) would extend faithful pictures to complex AME states such as those of Refs.~\cite{rather,bruzda2024}. For $d=3$ and $4$, local transformations of the permutation solutions of Sec.~\ref{sec:motivation} might yield genuine geomagic squares with non-congruent pieces. States of more parties would require higher-dimensional arrays.

\begin{acknowledgments}
The Center for Quantum-Enabled Computing project is carried out within the International Research Agendas programme of the Foundation for Polish Science co-financed by the European Union under the European Funds for Smart Economy 2021--2027 (FENG). The author is indebted to Maciej Bartula, Mykhailo Hontarenko, Palash Pandya, Saverio Pascazio, Suhail Ahmad Rather, Lee Sallows and Karol Życzkowski for many insightful discussions. The author acknowledges the use of ChatGPT (OpenAI, GPT-6 Astra Pro) for searching for (genuinely better) pictorial representations of AME states and performing exact computational checks, and of Claude (Anthropic, Claude Opus 5.5) for preparing the reproduction code and figures. The author takes responsibility for the final results and presentation.
\end{acknowledgments}

\appendix
\begin{table*}[t]
\centering\small
\setlength{\tabcolsep}{4pt}
\caption{Complete colored transversals. The weight is per color word and a dash means that only $\sigma_1$ is used.}\label{tab:transversals}
\begin{minipage}[t]{0.485\textwidth}
\centering
\begin{tabular}{@{}cc@{\quad}ccc@{\quad}c@{}}
\toprule $k$&$m$&$\tau$&$\sigma_1$&$\sigma_2$&$w$\\\midrule
0 & 0 & \texttt{251340} & \texttt{253104} & \texttt{-} & $1/4$ \\
0 & 1 & \texttt{341205} & \texttt{243150} & \texttt{-} & $1/4$ \\
0 & 2 & \texttt{430215} & \texttt{423150} & \texttt{-} & $1/4$ \\
0 & 3 & \texttt{520341} & \texttt{523104} & \texttt{-} & $1/4$ \\
1 & 0 & \texttt{253014} & \texttt{350241} & \texttt{-} & $1/4$ \\
1 & 1 & \texttt{342051} & \texttt{340215} & \texttt{-} & $1/4$ \\
1 & 2 & \texttt{432150} & \texttt{430215} & \texttt{-} & $1/4$ \\
1 & 3 & \texttt{523104} & \texttt{530241} & \texttt{-} & $1/4$ \\
2 & 0 & \texttt{035142} & \texttt{314025} & \texttt{-} & $1/4$ \\
2 & 1 & \texttt{124035} & \texttt{315042} & \texttt{-} & $1/4$ \\
2 & 2 & \texttt{215043} & \texttt{134025} & \texttt{-} & $1/4$ \\
2 & 3 & \texttt{304125} & \texttt{135042} & \texttt{-} & $1/4$ \\
3 & 0 & \texttt{031452} & \texttt{201534} & \texttt{-} & $1/4$ \\
3 & 1 & \texttt{120534} & \texttt{201453} & \texttt{-} & $1/4$ \\
3 & 2 & \texttt{210453} & \texttt{021534} & \texttt{-} & $1/4$ \\
3 & 3 & \texttt{301524} & \texttt{021453} & \texttt{-} & $1/4$ \\
\bottomrule
\end{tabular}
\end{minipage}
\hfill
\begin{minipage}[t]{0.485\textwidth}
\centering
\begin{tabular}{@{}cc@{\quad}ccc@{\quad}c@{}}
\toprule $k$&$m$&$\tau$&$\sigma_1$&$\sigma_2$&$w$\\\midrule
4 & 0 & \texttt{142530} & \texttt{415320} & \texttt{504321} & $1/16$ \\
4 & 1 & \texttt{145203} & \texttt{412503} & \texttt{502413} & $1/16$ \\
4 & 2 & \texttt{152403} & \texttt{415302} & \texttt{504312} & $1/16$ \\
4 & 3 & \texttt{154320} & \texttt{402531} & \texttt{512430} & $1/16$ \\
4 & 4 & \texttt{403512} & \texttt{054312} & \texttt{145302} & $1/16$ \\
4 & 5 & \texttt{405231} & \texttt{042531} & \texttt{152430} & $1/16$ \\
4 & 6 & \texttt{503421} & \texttt{054321} & \texttt{145320} & $1/16$ \\
4 & 7 & \texttt{504312} & \texttt{052413} & \texttt{142503} & $1/16$ \\
5 & 0 & \texttt{042513} & \texttt{415302} & \texttt{504312} & $1/16$ \\
5 & 1 & \texttt{045231} & \texttt{412530} & \texttt{502431} & $1/16$ \\
5 & 2 & \texttt{052431} & \texttt{415320} & \texttt{504321} & $1/16$ \\
5 & 3 & \texttt{054312} & \texttt{402513} & \texttt{512403} & $1/16$ \\
5 & 4 & \texttt{413520} & \texttt{054321} & \texttt{145320} & $1/16$ \\
5 & 5 & \texttt{415203} & \texttt{042513} & \texttt{152403} & $1/16$ \\
5 & 6 & \texttt{513402} & \texttt{054312} & \texttt{145302} & $1/16$ \\
5 & 7 & \texttt{514320} & \texttt{052431} & \texttt{142530} & $1/16$ \\
\bottomrule
\end{tabular}
\end{minipage}
\end{table*}

\section{Construction in Components}\label{app:construction-details}

\begin{figure}[ht!]
\centering
\includegraphics[width=\columnwidth]{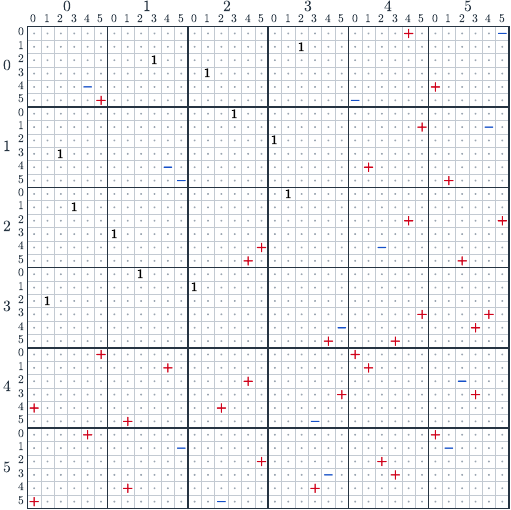}
\caption{The sparse real two-unitary matrix $U$. Rows are labeled by $ij$ and columns by $kl$. The symbols {\color{figzero}$\boldsymbol{\cdot}$}, $\mathbf{1}$, {\color{figplus}$\boldsymbol{+}$} and {\color{figminus}$\boldsymbol{-}$} denote $0$, $1$ and $\pm 1/\sqrt{2}$. Thick lines separate the $6\times6$ blocks of fixed $i$ and $k$. The geomagic figure encodes the locally equivalent $U^{\textrm{g}}$ of Eq.~\eqref{eq:local-maps}.}
\label{fig:U}
\end{figure}

The Pauli prescription in Section~\ref{sec:construction} can be read directly in matrix form. Figure~\ref{fig:U} displays $U_{6i+j,6k+l}=T_{ijkl}$. For instance, the entry $\Xi^{AD}_3=\textrm{X}$ of Eq.~\eqref{pair-table} determines
\begin{equation}
\begin{aligned}
T_{3,4,4,3}&=U_{22,27}=0=T_{3,5,5,3}=U_{23,33},\\
T_{3,4,5,3}&=U_{22,33}=1/\sqrt{2},\\
T_{3,5,4,3}&=U_{23,27}=1/\sqrt{2}.
\end{aligned}
\end{equation}
The twelve additional unit entries are the odd permutations of $(0,1,2,3)$:
\begin{equation}
\begin{gathered}
(0,1,3,2),\ (0,2,1,3),\ (0,3,2,1),\ (1,0,2,3),\\
(1,2,3,0),\ (1,3,0,2),\ (2,0,3,1),\ (2,1,0,3),\\
(2,3,1,0),\ (3,0,1,2),\ (3,1,2,0),\ (3,2,0,1).
\end{gathered}
\end{equation}
Thus the complete support contains $12$ coefficients equal to $1$, $34$ to $1/\sqrt2$ and $14$ to $-1/\sqrt2$.

There is also a Bell-pair interpretation. For $c<d$, define
\begin{equation}
\ket\Xi_{cd}=\frac{1}{\sqrt{2}}\sum_{u,v=0}^1\Xi_{uv}\ket{4+u}_c\ket{4+v}_d.
\end{equation}
The matrices $\textrm{I}$, $\textrm{Z}$, $\textrm{X}$ and $i\textrm{Y}$ give maximally entangled Bell states: $\Phi^+,\Phi^-,\Psi^+,\Psi^-$. Also, with $c$ and $d$ the complement of $a$ and $b$, one has
\begin{equation}
\ket\psi=\frac{1}{6}\sum_{q\ \textrm{odd}}\ket q+ \frac{1}{6}\sum_{ab}\sum_{x=0}^3\ket x_a\ket x_b\ket{\Xi_x^{ab}}_{cd}.
\end{equation}
Whenever two parties hold the same level $x\leqslant 3$, the other two form the prescribed Bell pair in levels $\{4,5\}$.

\section{Drawing and Reading the Geomagic Square}\label{app:geometry}

The procedure is technical in nature and can be summarized concisely in the following steps:

1. Placing the sectors: Construct $T$ as in Section~\ref{sec:construction} and obtain $T^{\textrm{g}}$ from Eq.~\eqref{eq:local-maps}. With unit circumradius, set vertices $V_s=(\cos(s_k\pi/3),\sin(s_k\pi/3))$ with $(s_0,\ldots,s_5)=(0,1,4,5,2,3)$. Sector $k$ has vertices $0$, $V_{s_k}$ and $V_{s_k+1}$ with subscripts taken modulo six. The tilings and the color balance do not depend on this placement of the sectors and the exact check shows that all $720$ placements also provide pairwise non-congruent pieces, both for the canonical and for the illustrated split~\cite{github}. Figure~\ref{fig:geomagic} uses $(0,1,4,5,2,3)$. An edge interval $[a,b]$ gives the triangle $0$, $(1-a)V_{s_k}+aV_{s_k+1}$ and $(1-b)V_{s_k}+bV_{s_k+1}$. Its area is equal to $b-a$ in units of one sector (one sixth of the hexagon). One should use the displayed row and column orders from Section~\ref{sec:geomagic}.

2. Subdivision and coloring: A word in Table~\ref{tab:transversals} lists a permutation's images at $i=0,\ldots,5$, including leading zeros. Matching $m$ of sector $k$, with permutation $\tau=\tau_{km}$, selects cells $(i,\tau(i))$ and color word $\sigma$ assigns color $l=\sigma(i)$. Each color word has weight $w$. The $32$ records yield $48$ weighted transversals, i.e., each transversal $\alpha$ in Eq.~\eqref{eq:geomagic-decomposition} is a matching $m$ together with one of its color words $\sigma$, so that $\tau_{k\alpha}=\tau_{km}$, $\sigma_{k\alpha}=\sigma$ and $w_{k\alpha}=w$.

All $36$ cell hexagons share the same subdivision of every sector and the matching of an interval determines the six cells in which its triangle is drawn. Along the outer edge of sector $k$, from $V_{s_k}$ to $V_{s_k+1}$, consecutive intervals belong to the matchings listed by the $k^{\textrm{th}}$ word:
\begin{equation}\label{eq:matching-words}
\begin{array}{c|l||c|l}
k&\text{matching word}&k&\text{matching word}\\\hline
0&\texttt{23310102}&3&\texttt{11302230}\\
1&\texttt{21303021}&4&\texttt{7632321056014475}\\
2&\texttt{32113020}&5&\texttt{6013617402535724}\\
\end{array}.
\end{equation}
Every matching occurs twice. If it has $q$ color words, one assigns its two occurrences arbitrary positive lengths with sum $qw$. Equal halves have length $1/8$ per occurrence for $k<4$ and $1/16$ for $k=4$ and $5$ (see Sec.~\ref{sec:two}), providing a canonical drawing. Figure~\ref{fig:geomagic} uses unequal splits and its exact coordinates are reproduction data in~\cite{github}.

Next, place each interval's triangle in the six cells selected by its matching. For one color word, use $\sigma_1$. For two, traverse that matching's occurrences in edge order, namely assign the first cumulative length $w$ to $\sigma_1$ and the remaining $w$ to $\sigma_2$, splitting an interval if necessary. Hatch a fragment if its coefficient $T^{\textrm{g}}_{ijkl}$ from Eq.~\eqref{eq:local-maps} is negative. These rules define the pieces for every allowed positive split.

3. Checking non-congruence: For each edge in cyclic order (sectors $0,1,4,5,2,3$), replace each letter $m$ of Eq.~\eqref{eq:matching-words} by $1$ if $j=\tau_{km}(i)$ and by $0$ otherwise. Count the blocks of consecutive ones on each edge separately. These six counts form the edge word of cell $(i,j)$. Its total number of lobes is the sum of these six counts minus the joins across adjacent occupied edge ends.

Inspection reveals $3$--$12$ lobes per piece and $32$ distinct edge words up to rotation and reflection. Only four pairs of pieces share an edge word: $(0,3)/(1,3)$, $(0,4)/(4,0)$, $(0,5)/(5,1)$ and $(2,4)/(2,5)$. In the first two pairs the lobe counts differ, $8$ versus $6$ and $10$ versus $11$. In the last two pairs the common words, $123400$ and $004420$, are left unchanged by no rotation or reflection except the identity, so a congruence would have to be the identity. This is excluded, since the first piece of each pair occupies a positive-length interval at the start of sector $4$'s outer edge and it is not the case for the second one. Hence all pieces are pairwise non-congruent. The test uses only the matching order and the positivity of the lengths, so it holds for every split which is allowed.

4. Reading the matrix: Use the central $6\times6$ array of Fig.~\ref{fig:geomagic}. Read $(i,j)$ from the cell labels: $k$ from the sector key and $l$ from the color legend. Let $\mathcal F_{ijkl}$ denote the set of fragments of color $l$ in sector $k$ of cell $(i,j)$ and define
\begin{equation}\label{eq:decode-area}
A_{ijkl}=\sum_{F\in\mathcal F_{ijkl}}\frac{\operatorname{area}(F)}{\operatorname{area}(\text{one sector})}.
\end{equation}
Instead of angular width one should use area (or fractional edge). All fragments of one coefficient share its hatching. The summation must be performed first, followed by taking the signed square root:
\begin{equation}\label{eq:decode-entry}
U^{\textrm{g}}_{6i+j,6k+l}=
\left\{\renewcommand{\arraystretch}{1.2}\begin{array}{@{}c@{\;:\;}l@{}}
0&A_{ijkl}=0,\\
-\sqrt{A_{ijkl}}&\text{hatched},\\
+\sqrt{A_{ijkl}}&\text{unhatched}.
\end{array}\right.
\end{equation}
For example, the two yellow--hatched fragments for $(i,j,k,l)=(0,4,0,4)$ have combined area $1/4$ in sector units, resulting in $U^{\textrm{g}}_{4,4}=-1/2$. The normalized state amplitude is the matrix entry divided by six.

Exact aggregate areas also follow directly from Table~\ref{tab:transversals}. Let $A=0$ and add $w$ to $A_{i,\tau(i),k,\sigma(i)}$ for each $i$ for every listed color word $\sigma$. Signs are read from hatching or calculated from Eq.~\eqref{eq:local-maps}. The original sparse matrix can be recovered by $U=(O_A\otimes O_B)^{\mathsf T}U^{\textrm{g}}(O_C\otimes\Id_6)$.

\interlinepenalty=10000
\bibliography{RAME46}
\end{document}